\documentclass[10pt,prd,preprintnumbers,floatfix,aps,nofootinbib,showpacs,twocolumn,superscriptaddress,noshowpacs]{revtex4} 

\makeatletter
\g@addto@macro\bfseries{\boldmath}
\makeatother

\usepackage{epsfig}
\usepackage{url}
\usepackage{hyperref}
\usepackage{soul}

\usepackage[normalem]{ulem} 

\usepackage{latexsym}
\usepackage{epsfig}
\usepackage{amsmath}
\usepackage{bm}
\usepackage{amssymb}
\usepackage{wasysym}
\usepackage{graphicx}
\usepackage{verbatim}
\usepackage{enumerate,mdwlist}
\usepackage[titletoc]{appendix}
\usepackage{amsfonts}
\usepackage{tikz} 
\usetikzlibrary{calc}
\usepackage{pgfplots}
\usepackage[export]{adjustbox}

\definecolor{DARKBLUE}{HTML}{00008b}
\hypersetup{colorlinks=true
,urlcolor=DARKBLUE
,anchorcolor=DARKBLUE
,citecolor=DARKBLUE
,filecolor=DARKBLUE
,linkcolor=DARKBLUE
,menucolor=DARKBLUE
,linktocpage=true
,pdfproducer=medialab
}

\usepackage{bbold}
\usepackage[normalem]{ulem}

\newcommand{\ie}{{i.e.~}}

\newcommand{\eg}{e.g.~}

\newcommand{\etc}{etc.}

\newcommand{\LCDM}{$\Lambda$CDM}

\let\oldsqrt\sqrt
\def\sqrt{\mathpalette\DHLhksqrt}
\def\DHLhksqrt#1#2{%
\setbox0=\hbox{$#1\oldsqrt{#2\,}$}\dimen0=\ht0
\advance\dimen0-0.2\ht0
\setbox2=\hbox{\vrule height\ht0 depth -\dimen0}%
{\box0\lower0.4pt\box2}}

\newcommand{\dd}{\mathrm{d}}
\newcommand{\ee}{e}

\newcommand{\umax}{\mathrm{max}}

\newcommand{\uNL}{\mathrm{NL}}

\newcommand{\fNL}{f_\uNL}

\newcommand{\beq}{\begin{equation}}
\newcommand{\eeq}{\end{equation}}
\newcommand{\bea}{\begin{equation}\begin{aligned}}
\newcommand{\eea}{\end{aligned}\end{equation}}

\newlength{\wsingfig}
\newlength{\wdblefig}
\newlength{\wquadfig}
\newlength{\wtriplefig}
\newcommand{\Eq}[1]{Eq.~(\ref{#1})}

\newcommand{\Fig}[1]{Fig.~{\ref{#1}}}

\newcommand{\Refc}[1]{Ref.~{\cite{#1}}}
\newcommand{\Refs}[1]{Refs.~{\cite{#1}}}

\definecolor{grey}{rgb}{0.4,0.4,0.4}
\definecolor{dullmagenta}{rgb}{0.4,0,0.4}
\definecolor{darkblue}{rgb}{0,0,0.4}
\definecolor{midblue}{rgb}{0,0,0.5}
\definecolor{midred}{rgb}{0.5,0,0}
\definecolor{orange}{rgb}{1,0.5,0}
\definecolor{lightbrown}{rgb}{0.75,0.5,0.25}
\definecolor{tan}{cmyk}{0.14,0.42,0.56,0}
\definecolor{djunglegreen}{cmyk}{0.99,0,0.52,0}
\definecolor{lightgreen}{rgb}{0,1,0}
\definecolor{olivegreen}{cmyk}{0.64,0,0.95,0.40}
\definecolor{midgreen}{rgb}{0.0,0.675,0.0}
\definecolor{darkgreen}{rgb}{0,0.5,0}

\newcommand{\s}{\sigma}
\newcommand{\sG}{\sigma_{\mathrm{G}}}

\usepackage[all]{xy} 
\usepackage{amsfonts}

\begin{document} 

\title{Massive galaxy clusters like ``El Gordo'' hint at primordial quantum diffusion}

\author{Jose Mar\'ia Ezquiaga}
\email{jose.ezquiaga@nbi.ku.dk}
\affiliation{Niels Bohr International Academy, Niels Bohr Institute, Blegdamsvej 17, DK-2100 Copenhagen, Denmark}
\affiliation{NASA Einstein fellow; Kavli Institute for Cosmological Physics and Enrico Fermi Institute, The University of Chicago, Chicago, IL 60637, USA}

\author{Juan Garc\'ia-Bellido}
\email{juan.garciabellido@uam.es}
\affiliation{Instituto de F\'isica Te\'orica UAM-CSIC, Universidad Aut\'onoma de Madrid, Cantoblanco,
Madrid, 28049 Spain}

\author{Vincent Vennin}
\email{vincent.vennin@ens.fr}
\affiliation{Laboratoire de Physique de l'\'Ecole Normale Sup\'erieure, ENS, Universit\'e PSL, CNRS, Sorbonne Universit\'e, Universit\'e Paris Cit\'e, F-75005 Paris, France}
\affiliation{Laboratoire Astroparticule et Cosmologie, CNRS Universit\'e de Paris, 10 rue Alice
Domon et L\'eonie Duquet, 75013 Paris, France}

\begin{abstract}
It is generally assumed within the standard cosmological model that initial density perturbations are Gaussian at all scales. However, primordial quantum diffusion unavoidably generates non-Gaussian, exponential tails in the distribution of inflationary perturbations. These exponential tails have direct consequences for the formation of collapsed structures in the universe, as has been studied in the context of primordial black holes. We show that these tails also affect the very-large-scale structures, making heavy clusters like ``El Gordo'', or large voids like the one associated with the cosmic microwave background cold spot, more probable. We compute the halo mass function and cluster abundance as a function of redshift in the presence of exponential tails. We find that quantum diffusion generically enlarges the number of heavy clusters and depletes subhalos, an effect that cannot be captured by the famed $\fNL$ corrections. These late-universe signatures could thus be fingerprints of quantum dynamics during inflation that should be incorporated in $N$-body simulations and checked against astrophysical data.
\end{abstract}

\date{\today}
 
\maketitle
 

\textbf{\emph{Introduction.}} 
\label{sec:Intro}
The standard cosmological model (\LCDM), provides an excellent fit to the high-precision astrophysical and cosmological observations, in particular the Cosmic Microwave Background (CMB), the Large-Scale Structure (LSS) of the universe and the relative abundance of light elements. 
Its three main ingredients are: (1) General Relativity and the Cosmological Principle, (2) a universe made of baryonic matter, dark matter, radiation and dark energy, and (3) quasi scale-invariant, Gaussian initial density fluctuations. 

However, hints for a few cracks start to emerge at different stages, in the form of moderate statistical tensions in parameter inference, e.g. the local expansion rate \cite{Freedman:2017yms,Verde:2019ivm}, 
or via the existence of ``extreme'' objects or outliers, which are more frequently observed than what \LCDM~predicts. Those may be associated with either an extremely low value of the density field (such as the Eridanus supervoid~\cite{Finelli:2014yha,Kovacs:2015hew}, which seems to have a direct connection with the CMB cold spot~\cite{Kovacs:2021wnc}), or with extremely large values of the density field (such as massive galaxy clusters like {\textit{El Gordo}}~\cite{Asencio:2020mqh} -- see however \Refc{Kim:2021bjb} for a recent smaller estimate of its mass -- and the presence of galaxies and Quasi-Stellar Objects at extremely high redshifts, where according to standard \LCDM~ there should not be any~\cite{Gonzalez:2012pk,Finkelstein:2013lfa}). 
In addition to the early structure formation issues, there are late time miss-matches at small scales such as 
the substructure problems~\cite{DelPopolo:2016emo}, the too-big-to-fail and the core-cusp problems~\cite{Newton:2017xqg}, which could be alleviated by incorporating baryonic physics~\cite{Zavala:2019gpq}.

While most attempts to reconcile those potential issues  
focus on relaxing either the first or the second assumption  mentioned above (\ie modifying the laws of gravity, or invoking the existence of additional components in the universe), a natural strategy to accommodate the existence of extreme objects within the \LCDM~paradigm would be to question the third assumption, namely the Gaussianity of the primordial density fluctuations. 
The reason is 
twofold: experimentally, there are more extreme objects than what Gaussian tails suggest, pointing toward the existence of heavier tails; theoretically, the typical mechanisms producing primordial cosmological perturbations anyway lead to non-Gaussian tails.  

In the early universe indeed, vacuum quantum fluctuations are amplified by gravitational instability and stretched to large distances, giving rise to classical fluctuations in the density field, that later collapse into cosmological structures.\footnote{This mechanism is mostly studied in the context of inflation, but it also operates in most of its alternatives~\cite{Brandenberger:2009jq}.} 
At leading order in cosmological perturbation theory, it gives rise to Gaussian perturbations, in good agreement with CMB measurements~\cite{Akrami:2019izv}. 
However, the CMB gives access to large scales only and leaves small scales mostly unconstrained. Moreover, even at large scales, they restrict the statistics of the most likely fluctuations only, \ie they reconstruct only the neighbourhood of the maximum of the underlying distribution functions, and say little about their tails.

Nonetheless, beyond linear order, those tails are expected to be non-Gaussian.
The difficulty when characterising the statistics of those tails is that they require non-perturbative techniques. 
Perturbative approaches, such as calculations of the bi- or tri-spectrum, and $\fNL$-like parametrizations, are tailored to describe small deviations from Gaussianity around the maximum, not accounting for the tails.

\textbf{\emph{Quantum diffusion and non-Gaussian tails.}} 
Recently, non-perturbative techniques have been developed to study how quantum diffusion, the presence of which is inevitable in scenarios where cosmological perturbations have a quantum origin, modifies the expansion dynamics of the universe and thus affects the statistics of density fluctuations. 
This can be done by combining three approaches to describe the dynamics of super-Hubble degrees of freedom. 
First, the separate-universe picture~\cite{Wands:2000dp, Khalatnikov:2002kn, Lyth:2003im, Lyth:2004gb} (which is valid beyond slow roll~\cite{Pattison:2019hef, Artigas:2021zdk}), according to which spatial gradients can be neglected on super-Hubble scales, and each spatial point evolves independently along the dynamics of an unperturbed universe. 
Second, stochastic inflation~\cite{Starobinsky:1982ee, Starobinsky:1986fx}, in which quantum fluctuations act as a stochastic noise on the classical, background evolution of each of these separate universes. 
Third, the $\delta N$ formalism~\cite{Sasaki:1995aw, Sasaki:1998ug, Lyth:2004gb, Lyth:2005fi}, which states that, in each of these separate universes, the local fluctuation in the amount of expansion realized between an initial flat hypersurface and a final hypersurface of uniform energy density is nothing but the curvature perturbation.
This gives rise to the stochastic-$\delta N$ formalism~\cite{Enqvist:2008kt, Fujita:2013cna, Fujita:2014tja, Vennin:2015hra, Vennin:2016wnk, Firouzjahi:2018vet}, which provides a non-perturbative scheme to compute the statistics of curvature perturbations on super-Hubble scales.
These methods now extend to the calculation of the density contrast and the compaction function~\cite{Tada:2021zzj}.

While these techniques recover quasi-Gaussian distributions close to their maximum, with $\fNL$-type corrections, they also reveal the existence of systematic exponential tails~\cite{Pattison:2017mbe, Ezquiaga:2019ftu, Vennin:2020kng, Figueroa:2020jkf, Ando:2020fjm, Pattison:2021oen, Rigopoulos:2021nhv, Tada:2021zzj}, which strongly deviate from the Gaussian profile (such heavy tails were also found in \Refs{Panagopoulos:2019ail, Achucarro:2021pdh, Kuhnel:2021yic, Cai:2022erk} using different methods).
More precisely, the distribution function of the first-passage time $\mathcal{N}$ can be expanded as $P(\mathcal{N}) = \sum_{n\geq 0} a_n(\bm{\Phi}) \ee^{-\Lambda_n \mathcal{N}}$, in which $\Lambda_n$ are the eigenvalues of the adjoint Fokker-Planck operator associated with the stochastic problem under consideration, and $ a_n(\bm{\Phi})$ are coefficients that depend on the initial configuration in field space (here denoted as $\bm{\Phi}$). Far on the tail, the smallest eigen-value dominates, $P(\mathcal{N})\propto \ee^{-\Lambda_0 \mathcal{N}}$, which implies that large perturbations are much more likely than what a Gaussian behavior, $P_\mathrm{G}\propto \ee^{-\Lambda\mathcal{N}^2}$, would suggest. 
In practice, these exponential tails are more important in models where quantum diffusion dominates at some stages of the inflationary dynamics (leading to smaller values of $\Lambda_0$, hence heavier tails). Depending on the time at which this happens, they affect structures at different scales. 
If the fluctuations are large enough, they may even collapse into black holes upon horizon re-entry after inflation. This is why non-Gaussian tails have been mostly studied in the context of primordial-black-hole production (see \eg \Refs{Clesse:2015wea, Kawasaki:2015ppx,Pattison:2017mbe, Ezquiaga:2018gbw, Biagetti:2018pjj,Figueroa:2020jkf,Pattison:2021oen,Figueroa:2021zah,Kitajima:2021fpq,Tada:2021zzj}).

Nonetheless, as we argue here, these heavy tails may also play a key role in the formation of the LSS and point toward potential solutions to some of the problems of \LCDM. 
Importantly, while the parameters $a_n$ and $\Lambda_n$ depend on the details of the model under consideration, the existence of these exponential tails is ubiquitous and arises in any model where quantum diffusion is at play. In this sense, they are already embedded in the \LCDM~scenario. Therefore, our approach does not rely on extending \LCDM~to solve the above-mentioned issues: our goal is rather to point out that \LCDM~may already contain the ingredients needed to explain those ``anomalous'' observations, provided we carefully compute the primordial statistics beyond the perturbative level. 

Heavy tails in the form of lognormal distributions are already known to develop on sub-Hubble scales after inflation, due to gravitational collapse~\cite{Coles:1991if, Taruya:2002vy,Hilbert:2011xq}. However, the effect we are considering here is different: it leads to \emph{primordial} heavy tails, which are present even before Hubble re-entry.

\

\textbf{\emph{Exponential tails in the primordial statistics of perturbations.}} 
The details of the stochastic distribution associated with primordial perturbations depend on the specifics of the inflationary model (the number of fields, their potential, their kinetic coupling \etc). 
In order to describe the amplitude of fluctuations coarse-grained at a certain scale, one has to convolve the first-passage time distributions against backward distributions of the field value~\cite{Ando:2020fjm,Tada:2021zzj}. 
Moreover, one must account for the non-linear mapping between the curvature perturbation and the density contrast~\cite{Musco:2018rwt}, which further modifies distribution functions and can also introduce heavy tails~\cite{Biagetti:2021eep,DeLuca:2022rfz}. 
In this work, we do not aim at deriving predictions for specific models, but rather wish to explore generic consequences arising from the presence of heavy tails. This is why, in practice, we consider two normalized templates for the distribution function of the density contrast in comoving threading $\delta$,
\bea \label{eq:elliptic}
P_2(\delta_k)  &= -\frac{\pi}{2\mu^2}\vartheta_2'\left(\frac{\pi \alpha_k}{2},\ee^{-\frac{\pi^2}{\mu^2}\mathcal{D}_k}\right)\, ,\\
P_4(\delta_k) &=
\frac{\pi}{2\mu^2\alpha_k}\vartheta_4'\left(\frac{\pi \alpha_k}{2},\ee^{-\frac{\pi^2}{\mu^2}\mathcal{D}_k}\right)\, .
\eea 
In these expressions, $\delta_k$ denotes the Fourier mode of the density contrast, related to the positive variable $\mathcal{D}_k$ through the relation $\delta_k=\mathcal{D}_k-\langle \mathcal{D}_k \rangle$ where the mean value is taken with respect to the distribution function in question. These distributions depend on two parameters, $\alpha_k$ and $\mu$, the latter being scale independent to reflect the fact that the eigenvalues $\Lambda_n$ do not depend on the field configuration, hence on the scale~\cite{Ezquiaga:2019ftu}. Finally, $\vartheta'_2$ and $\vartheta'_4$ are the derivatives of the elliptic theta functions of the second and fourth kind respectively~\cite{nist}. In what follows, they are refereed to as the ``elliptic 2'' and ``elliptic 4'' templates respectively. Such functions are often found in toy models of quantum diffusion~\cite{Pattison:2017mbe, Ezquiaga:2019ftu}. 

\begin{figure}[t]
\includegraphics[width =\columnwidth]{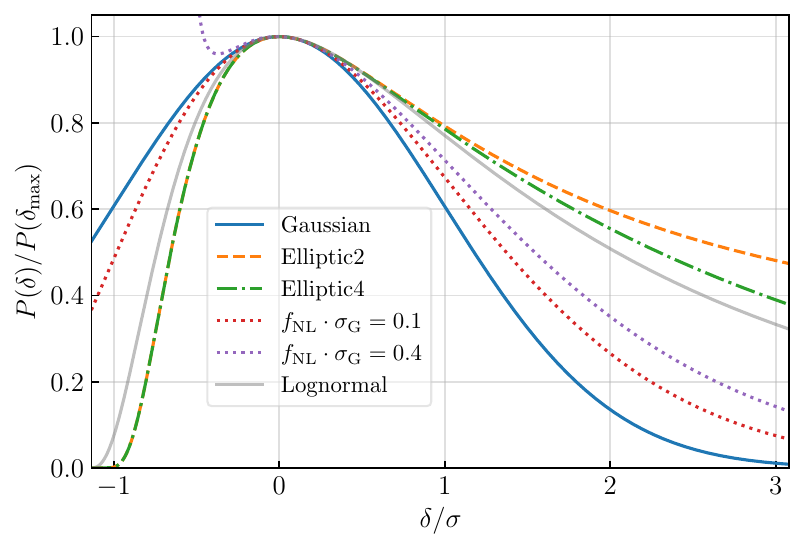}
\caption{Gaussian, elliptic, local-$\fNL$ and lognormal distributions, as a function of $\delta/\sigma$ where $\sigma$ is the standard deviation of the corresponding distribution. 
The free parameters of those distributions are set such that they all share the same value of $\sigma$ around the maximum and are given by $\alpha=0.5$, $\mu=\pi$ and $\sigma=\sqrt{2}\alpha^2$, with the same $\alpha$ and $\mu$ for both elliptic functions. 
See Supplemental Material for how to match the shape around the maximum.}
\label{fig:comparison_pdf}
\end{figure}

The two distributions are displayed in \Fig{fig:comparison_pdf} as a function of $\delta/\sigma$ where hereafter $\sigma$ denotes the standard deviation of the distribution under consideration, and where they are compared with a Gaussian distribution, a local $\fNL$ distribution and a lognormal distribution. The free parameters of those distributions are set such that they are maximal at the same location and all share the same value of $\sigma$, see the Supplemental Material where this procedure is further detailed. Both elliptic profiles are endowed with a heavier upper tail, and with a lighter lower tail, than the Gaussian fit. 

The local $\fNL$ parametrisation is defined as
\bea
\label{eq:fNL:def}
\delta(x) = \delta_{\mathrm{G}}(x) + \frac{3}{5}\,\fNL \Big[\delta_{\mathrm{G}}^2(x) - \sG^2\Big]\, ,
\eea
where $\delta_{\mathrm{G}}$ has a Gaussian distribution function centered at zero and with dispersion $\sG\equiv\langle\delta_{\mathrm{G}}^2\rangle^{1/2}$. 
From this expression, one can show that
\begin{equation}
   \label{eq:P:NL}
P_{\mathrm{NL}}(\delta) =
\frac{1}{\sqrt{2\pi\sG^2\Delta}}\left[e^{-\frac{25(\sqrt{\Delta}-1)^2}{72 \fNL^2 \sG^2}}+e^{-\frac{25(\sqrt{\Delta}+1)^2}{72 \fNL^2 \sG^2}}\right]\,,
\end{equation}
where $\Delta(\delta) = 1+ 12/5\,\fNL\delta + 36/25\,\fNL^2\sG^2$.
As shown in \Fig{fig:comparison_pdf}, 
although $\fNL$ correctly describes the non-Gaussian corrections around the maximum, it fails to capture the highly non-Gaussian tails. 
Similarly, since $P_{\mathrm{NL}}$ diverges when $\delta$ approaches $-3 \fNL \sigma^2/5-5/(12\fNL)$, it cannot properly describe the small-$\delta$ statistics.\footnote{The local $\fNL$ in Eq.~(\ref{eq:fNL:def}) is assumed to be constant, and thus cannot reproduce the scale-dependence of the amplitude of the tails that comes from the fact that quantum diffusion has a different effect at different points of the inflationary evolution. Other formulations of primordial non-Gaussianity take into account the ``scale-dependence" of $\fNL$ through calculations of the full bispectrum, but they would still be perturbative.}
Interestingly, the elliptic functions are more similar to a lognormal distribution than the perturbative $\fNL$ approximation.

\

\textbf{\emph{Implications for the Large-Scale Structure.}} 
\begin{figure*}[t]
\includegraphics[width = 0.9\textwidth]{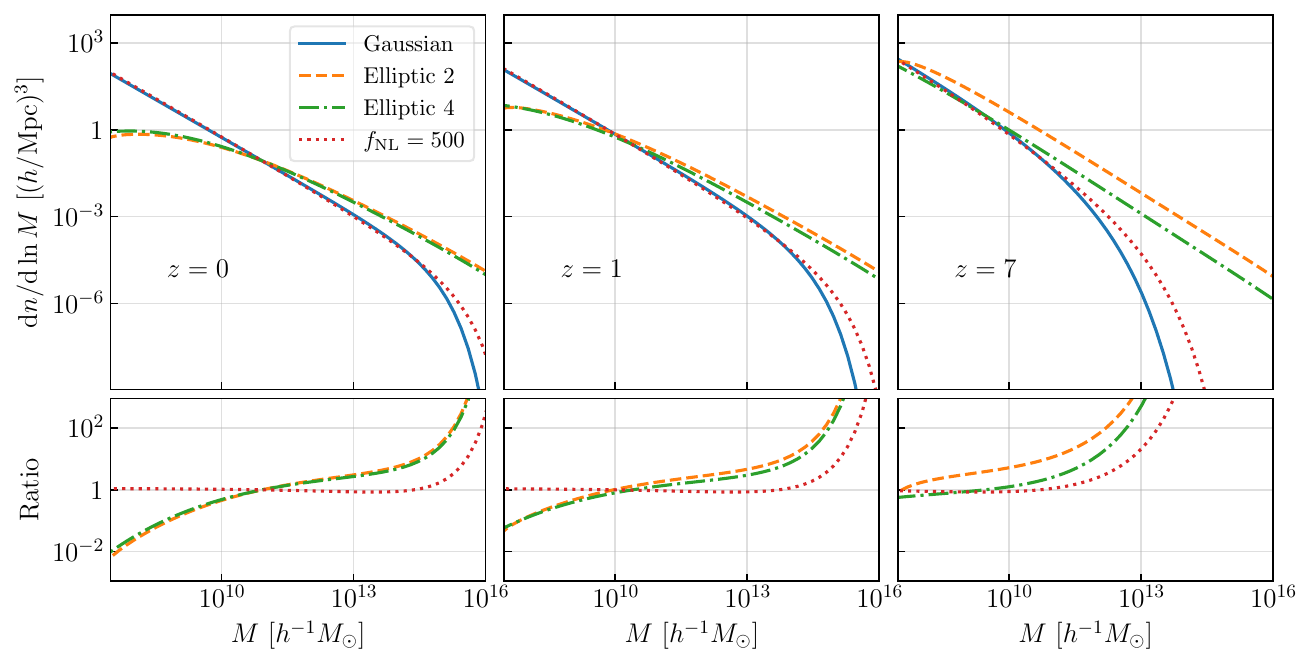}
\caption{Halo mass function (\ie differential number of halos per comoving volume) obtained from different distributions for the primordial density perturbations: Gaussian, elliptic 2 and 4, and local $\fNL$ (where $\fNL$ is fixed at the last scattering surface). 
Each column corresponds to the HMF at a different redshift. The bottom panels show the ratio between the HMF and the Gaussian result. Quantum diffusion affects both low- and high-mass ends of the HMF and become more significant at higher redshifts, making their signatures distinguishable from perturbative non-Gaussianities ($\fNL$). 
The normalization is fixed to match the Gaussian at $M=10^{11} h^{-1}M_\odot$ and $z=0$, where $h$ is the dimensionless Hubble constant, $h=H_0/(100$ km/s/Mpc).
}
\label{fig:HMF}
\end{figure*}
The simplest statistics to be extracted from the primordial density fluctuations is the one-point function, \ie the number of collapsed objects. 
Following the Press-Schechter formalism \cite{Press:1973iz}, this is given by the probability that $\delta$ is above a given threshold $\delta_{\mathrm{c}}$,
$\beta=P(\delta>\delta_{\mathrm{c}})=2\int_{\delta_{\mathrm{c}}}^\infty P(\delta)\dd\delta$. 
$\delta_{\mathrm{c}}$ depends on the time of re-entry of the fluctuations and has been extensively explored in the literature \cite{Harada:2013epa,Young:2014ana,Germani:2018jgr,Yoo:2018kvb}. 
For our purposes it will be enough to fix it to $\delta_{\mathrm{c}}=1.68$, as predicted by linear theory of spherical collapse \cite{1980lssu.book.....P}. 
From \Fig{fig:comparison_pdf}, it is clear that as $\nu\equiv\delta_{\mathrm{c}}/\sigma$ increases, 
the number of collapsed objects is larger in heavy-tailed models than in the Gaussian case.

More precisely, let us study how structures 
distribute across different masses. 
This can be achieved with the Halo Mass Function (HMF), defined from the mass fraction $\beta$ as
\bea\label{eq:HMF}
\frac{\dd n}{\dd\ln M} = \frac{\rho_{\mathrm{m}}}{M}\frac{\dd\beta}{\dd\ln M}=\frac{\rho_{\mathrm{m}}}{M}\,
\frac{\dd\ln\s^{-1}}{\dd\ln M}\,\nu\,\beta'(\nu)\,,
\eea
where $M$ is the mass of the halo, $\rho_{\mathrm{m}}$ is the energy density of matter and a prime denotes derivation with respect to $\nu$. Previous works have proposed to test $\fNL$ with the HMF, see \eg \cite{Matarrese:2000iz,Dalal:2007cu,LoVerde:2011iz,Yokoyama:2011sy,MoradinezhadDizgah:2020whw}. Here we extend those results by exploring initial density perturbations with non-Gaussian tails. 

Since present observations show a good agreement with the Gaussian hypothesis at galactic scales, we tune the free parameters of all considered distributions such that they peak at the same value and share the same standard deviation (see Supplemental Material for further details). As a consequence, the only difference in \Eq{eq:HMF} comes from the term $\beta'(\nu)$. In a Gaussian distribution, one has $\beta_{\mathrm{G}}'(\nu)=-2\ee^{-\nu^2/2}\sqrt{\pi}$, and similar expressions can be obtained for the other distributions.

Let us now study the redshift evolution of the number of halos. We can describe the HMF (\ref{eq:HMF}) as a function of redshift by writing $\rho_{\mathrm{m}}(a) = \Omega_{\mathrm{m}}(a)\,\rho_{\mathrm{c}}$ with $\Omega_{\mathrm{m}}(a) = \Omega_{\mathrm{m}}/a^3$ and $\s(a) = \s(1)\,D(a)$, with the growth function $D(a)=\delta(a)/\delta(1)$ given by~\cite{Buenobelloso:2011sja}
$D(a) \propto a\times{}_2F_1\left[(w-1)/2w,\,-1/3w,\,(6w-5)/6w,\,1-1/\Omega_{\mathrm{m}}(a)\right]\,,
$ where $w$ is the equation-of-state of Dark Energy (set to) $w=-1$ and $a$ is the scale factor.
By comparing the HMF at different redshifts with the abundance of massive clusters, we can estimate whether \eg ``El Gordo" is a typical cluster or not, at a given redshift.

Our main results are presented in \Fig{fig:HMF}, where we display the HMF for the four distributions under consideration at three redshifts, $z=0,1,7$. The bottom panels display the ratio of the HMF with respect to the Gaussian case, with the normalization fixed to the Gaussian at $M=10^{11} h^{-1}M_\odot$. 
At $z=0$ we observe two main effects from the exponential tails: an increase in the number of clusters ($M>10^{13}M_\odot$) and a decrease in the number of substructures ($M<10^{9}M_\odot$). Interestingly, the $\fNL$ reconstruction can partially mimic the increase in clusters, but not the decrease in substructures. This makes the predictions of quantum diffusion falsifiable. 
Equally important, the predictions of the elliptic HMF can potentially alleviate the shortcomings of \LCDM. 
Moreover, we also observe that the redshift evolution of the HMF is a key discriminator of the nature of the primordial perturbations. 
An elliptic HMF predicts that more massive objects formed earlier, in agreement with the recent detection of massive, high-redshift objects (see \eg \cite{Eilers:2021mdi} for a recent census of the age of young quasars).

In addition to the number of halos per unit mass, it is interesting to compute the number of clusters as a function of redshift. This can be probed directly for example with CMB data using the Sunyaev-Zeldovich (SZ) effect~\cite{Planck:2015koh,SPT:2019hnt,ACT:2020lcv}. 
Our results are presented in \Fig{fig:n_clusters}, focusing on clusters with {$M>10^{15}M_\odot$} (see Supplemental Material for the detailed calculation). One can clearly see that, already beyond $z\sim1$, the number of clusters is much enhanced when initial perturbations have heavy tails. 
This, again, shows the potential of this method to constrain the very early universe physics.

\begin{figure}[t]
\includegraphics[width =\columnwidth]{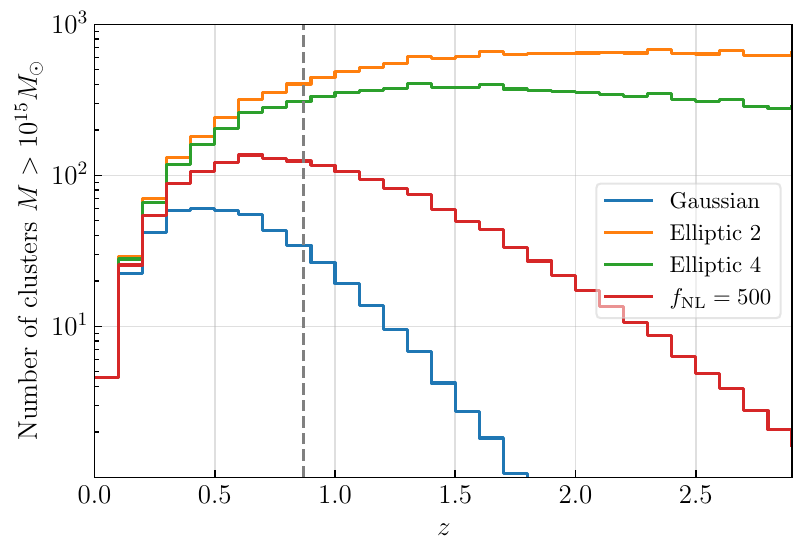}
\caption{Number of clusters with mass larger than $10^{15}M_\odot$ {in redshift bins of $\Delta z =0.1$.} 
as a function of redshift for the Gaussian, elliptic 2 and 4, and $\fNL$ distributions. The normalization is fixed to the Gaussian case at $z=0$. 
We highlight with a vertical dashed line the redshift of El Gordo ($z=0.87$).}
\label{fig:n_clusters}
\end{figure}

Observationally, in most cases we do not have direct access to the HMF, but rather to the amount of luminous matter. One thus needs to take into account the astrophysical systematics connecting these two. 
Recently, constraints on $\fNL$ have been derived using UV galaxy luminosity functions that marginalize over those systematics \cite{Sabti:2020ser}. 
A natural extension of this work would thus be to constrain the heavy tails from quantum diffusion with these data. 
Moreover, the HMF at sub-galactic scales could be probed analyzing the strong lensing rates and magnifications \cite{Gilman:2021gkj}.

\

\textbf{\emph{Future prospects.}} 
\LCDM~relies on the assumption of Gaussian initial conditions. 
Although CMB observations tightly constrain the amount of non-Gaussianities at large scales, little is known about the primordial fluctuations at smaller scales. 
Several processes in the early universe could lead to non-Gaussian distributions. 
Notably, an inevitable exponential tail arises due to quantum diffusion during inflation. 
In this work we have studied the imprints these heavy tails leave in the number of halos and their mass function. 
We have found that they enhance the number of heavy clusters and deplete the number of sub-halos, and that this difference with respect to the standard Gaussian initial conditions becomes more important at high redshift, depending on the strength of quantum diffusion. This could be compared with current SZ catalogs (e.g. Fig. 18 in \cite{ACT:2020lcv}) that did not find clusters of $M>10^{16}M_\odot$ and has a redshift distribution peaking at $z<1$. However, there are outstanding clusters like El Gordo with $M\sim3\times10^{15}M_\odot$ at $z=0.87$~\cite{Asencio:2020mqh}, and deep voids like the Eridanus supervoid~\cite{Kovacs:2021wnc}, and many more should soon be discovered with the James Webb Space Telescope (JWST)~\cite{Gardner:2006ky}, Euclid~\cite{Amendola:2016saw} and the Vera Rubin Observatory (LSST) \cite{LSST:2008ijt}. 

Let us note that the effect of quantum diffusion is similar to having a lognormal initial density distribution. Such lognormal profiles are indeed typically obtained from Gaussian initial conditions due to non-linear gravitational collapse~\cite{Coles:1991if}. In fact, to speed up the computation, many $N$-body codes start their evolution with an already lognormal distribution \cite{Agrawal:2017khv,Ramirez-Perez:2021cpq}. Here we find that non-linear growth occurs at much earlier times, as soon as non-Gaussian tails are present. This leaves specific features in the redshift-dependence of the statistics: for high-redshift galaxies we expect highly non-Gaussian statistics, much beyond what would be expected from non-linear gravitational collapse in such a short time.

Using the HMF to probe primordial universe physics requires further developments on various fronts.
From the observational side, we need to understand the systematics behind high-mass supergalactic structures at low and high redshifts, which Euclid~\cite{Amendola:2016saw} and JWST~\cite{Gardner:2006ky} observations may help to alleviate, as well as issues with baryonic physics and Halo Occupation Distributions. The HMF is sensitive to the one-point statistics of the density field only, but it would also be interesting to consider observables probing non-Gaussianities in higher correlators. From the theoretical side, we need to implement realistic physical models in our pipeline, and go beyond the simple phenomenological prescription adopted here (although this is not expected to alter our qualitative conclusions, it matters for quantitative details). 
Finally, on the numerical side, it would be necessary to run $N$-body simulations with non-Gaussian initial conditions of the type described above.

Altogether we have demonstrated the impact of quantum diffusion on the LSS and how the HMF and cluster abundances can probe early-universe physics. More importantly, we have shown that, within the standard cosmological model itself, quantum diffusion is inevitable during inflation, and some of the current tensions can be alleviated thanks to the non-Gaussian nature of the tails of primordial perturbations.

\begin{acknowledgments}
JGB acknowledges funding from the Research Project PGC2018-094773-B-C32 (MINECO-FEDER) and the Centro de Excelencia Severo Ochoa Program CEX2020-001007-S. 
JME is supported by the European Union's Horizon 2020 research and innovation program under the Marie Sklodowska-Curie grant agreement No. 847523 INTERACTIONS, and by VILLUM FONDEN (grant no. 53101 and 37766). 
He was also supported by NASA through the NASA Hubble Fellowship grant HST-HF2-51435.001-A awarded by the Space Telescope Science Institute, which is operated by the Association of Universities for Research in Astronomy, Inc., for NASA, under contract NAS5-26555; and by the Kavli Institute for Cosmological Physics through an endowment from the Kavli Foundation and its founder Fred Kavli.
\end{acknowledgments}


\section*{Supplemental Material}

In this supplemental material, we discuss how the elliptic templates can be approximated by other profiles, and where such approximations are expected to be valid. 

\section{Local $\fNL$ distribution} \label{app:elliptic}
Let us consider the elliptic profiles given in Eq.~(1) of the main text, and see how an effective $\fNL$ distribution, see Eq.~(3) of the main text, can be associated to them. 
\subsection{Matching the first moments}
A first approach consists in arranging the parameters of the $\fNL$ distribution such that the three first moments of both profiles coincide. The first moments of the $\fNL$ distribution can be calculated either from evaluating the expectation value of Eq.~(2) taken to some integer power, or from integrating Eq.~(3) of the main text directly. For the three first moments, one obtains 
\bea
\left\langle \delta \right\rangle_{\mathrm{NL}} =& 0\, ,\\
\left\langle \delta^2 \right\rangle_{\mathrm{NL}} =& \sG^2\left(1+\frac{18}{25} \fNL^2\sG^2\right)\, , \\
\left\langle \delta^3 \right\rangle_{\mathrm{NL}} =& \frac{18}{5}\fNL \sG^4\left(1+\frac{12}{25}\fNL^2\sG^2\right)\, .
\eea 
These relations can be inverted to obtain $\sG^2$ and $\fNL$ in terms of $\langle \delta^2\rangle$ and $\langle\delta^3\rangle$, leading to 
\bea 
\label{eq:sigma:fnl:NLdistrib}
\sG^2 = & \left\langle \delta^2 \right\rangle_{\mathrm{NL}}\left(X+\frac{1}{X}-1\right)\\
\fNL = & \frac{5}{3\sqrt{2\left\langle \delta^2 \right\rangle_{\mathrm{NL}}}}\frac{\sqrt{2-X-\frac{1}{X}}}{\left\vert {X+\frac{1}{X}-1}\right\vert}\mathrm{sign}\left(\left\langle \delta^3 \right\rangle_{\mathrm{NL}}\right)
\eea 
where 
\bea 
X = \left[1+\frac{\left\langle \delta^3 \right\rangle_{\mathrm{NL}}^2}{4\left\langle \delta^2 \right\rangle_{\mathrm{NL}}^3}\left(\sqrt{1-8\frac{\left\langle \delta^2 \right\rangle_{\mathrm{NL}}^3}{\left\langle \delta^3 \right\rangle_{\mathrm{NL}}^2}}-1\right)\right]^{1/3}\, .
\eea 
One may note that, when $8\left\langle \delta^2 \right\rangle_{\mathrm{NL}}^3>\left\langle \delta^3 \right\rangle_{\mathrm{NL}}^2$, $X$ is complex, but in that case one can readily check that $\vert X \vert =1$, hence $X+1/X$ is real.

From the three first moments of a given distribution, one can thus extract the corresponding $\sG^2$ and $\fNL$ parameters, hence the $\fNL$ distribution that reproduces those moments. 

\subsection{Matching the behaviour around the maximum}

The above procedure provides a good fit around the mean value of the reference distribution. However, elliptic profiles are such that the location of the mean and the location of the maximum are substantially different, and in practice it is more efficient to use the $\fNL$ distribution that best describes the behaviour around the maximum of the PDF. 

In the regime where $\Delta\gg (\sG \fNL)^4$, the second branch in Eq.~(3) of the main text can be neglected, and one can approximate
\bea
\label{eq:PNL:approx}
P_{\mathrm{NL}}(\xi) \simeq \frac{1}{\sqrt{2\pi\Delta}}
\exp\left[-\frac{(\sqrt{\Delta}-1)^2}{2\bar\sigma^2}\right] ,
\eea
where we have defined $\xi=\delta/\sG$ and $\bar\s = 6\fNL\sG/5$. The maximum of this distribution is at
$2\sqrt{\Delta_{\rm max}}=1+\sqrt{1-4\bar\sigma^2}\,,$ hence
\bea 
\xi_{\mathrm{max}} = \frac{\Delta_{\mathrm{max}} -1 - \bar\sigma^2}{2\bar\sigma}\,.
\eea 
This allows one to fix the $\fNL$ parameter such that the $\fNL$ distribution peaks at a given value of $\xi$. In Fig.~1 of the main text, we apply this procedure and display several $\fNL$ distributions (corresponding to several values of $\sG$) that share the same maximum location with the elliptic profiles. In order to accommodate the heavy tail, one can see that a large value of $\fNL \sigma$ needs to be used. 
However, when increasing $\fNL \sG$, the agreement at $\xi<\xi_{\rm max}$ becomes worse, and the point where the $\fNL$ distribution diverges gets dangerously too close to the maximum of the distribution. From this, one concludes that there is no $\fNL$ distribution that provides a reliable approximation of the elliptic profile both close to its maximum and along its tail.

In order to fix $\sG$, one can further expand \Eq{eq:PNL:approx} around $\xi=\xi_{\rm max}$, and one finds
\bea
P_{\mathrm{NL}}(\xi) &= P_{\rm NL}(\xi_{\umax}) \left[1 - 
\frac{(\xi-\xi_{\umax})^2}{2\sigma^2_{\umax}} + \dots
\right]
\eea
where
\bea
\sigma_{\mathrm{max}} &= \frac{\Big(1+\sqrt{1-4\bar\sigma^2}\Big)^2}{2\sqrt{2\Big(1-4\bar\sigma^2+\sqrt{1-4\bar\sigma^2}\Big)}}\, .
\eea
This can be used to match the curvature of the PDF around its maximum with a given reference distribution. It leads to the $\fNL$ profile with $\fNL\sigma_\mathrm{G}=0.1$ as plotted in Fig.~1 of the main text.

\section{Gaussian and lognormal distributions}

Similarly to what was done above, the elliptic profiles [see Eq.~(1) of the main text] can be approximated by a lognormal (LN) or a Gaussian (G) distribution,
\bea
P(x) = A\ {\rm LN}(x,\,\rho,\,\sigma) 
= A\,e^{-\frac{\sigma^2}{2}}\ {\rm G}(x,\,\rho,\,{\sigma_\mathrm{G}})\,,
\eea
where $x=\pi^2\mathcal{D} /\mu^2$,
\bea
{\rm LN}(x,\,\rho,\,\sigma) &= \frac{1}{\rho\,\sigma\sqrt{2\pi}}\,
\exp\left[-\frac{\ln(x/\rho)^2}{2\sigma^2}-\frac{\sigma^2}{2}\right] \nonumber \\[2mm]
{\rm G}(x,\,\rho,\,\sG) &= \frac{1}{\sG\sqrt{2\pi}}\,
\exp\left[-\frac{(x-\rho)^2}{2\sG^2}\right]
\eea
and $\sigma_{\mathrm{G}}=\rho\,\sigma$. In these parametrizations, $\rho$ stands for the value of $x$ where the PDF is maximal. It is the solution of the equations 
\bea\label{eq:series:2}
\sum_{n=0}^\infty  (2n+1)^3\,e^{-\rho\,n(n+1)}\sin\left[(2n+1)\frac{\pi}{2}\alpha\right] = 0\,,
\eea
for the elliptic 2 distribution, and 
\bea\label{eq:series:4}
\sum_{n=0}^\infty (-1)^n\, n^3\,e^{-\rho^2}\sin\left(n\pi\alpha\right) = 0
\eea
for the elliptic 4 distribution. When $\alpha$ is small, those equations have approximate solutions $\rho_2(\alpha)=\pi^2\alpha^2/6$ for the elliptic 2 distribution and $\rho_4(\alpha)=\pi^2(1-\alpha)^2/6$ for the elliptic 4 distribution (these approximations turn out to be reliable up until $\alpha\simeq 0.6$), otherwise those equations have to be solved numerically.
From here, the height of the elliptic distributions at their maxima can be inferred,
$P_{2/4}^\umax(\alpha) = P_{2/4}\left[\rho_{2/4}(\alpha),\alpha\right]$,
and equated with $A \ee^{-\sigma^2/2}/(\sigma_{\mathrm{G}}\sqrt{2\pi})$. 

The curvature of the elliptic distributions around their maximum can also be computed according to
\bea 
\sigma_{\mathrm{G}}^{(2)}(\alpha) = \left[-\left.\frac{\partial^2\ln P_2}{\partial x^2}\right\vert_{x=\rho_2(\alpha)}\right]^{-1/2}\, ,\\
\sigma_{\mathrm{G}}^{(4)}(\alpha) = \left[-\left.\frac{\partial^2\ln P_4}{\partial x^2}\right\vert_{x=\rho_4(\alpha)}\right]^{-1/2}\, .
\eea 
In the same small-$\alpha$ limit as above, they boil down to $\sigma_{\mathrm{G}}^{(2)}(\alpha)\simeq \sqrt{2}\alpha^2$ and $\sigma_{\mathrm{G}}^{(2)}(\alpha)\simeq \sqrt{2}(1-\alpha)^2$. This allows one to set the value of $\rho$, $\sigma_{\mathrm{G}}$ (hence $\sigma$) and $A$.

This procedure is performed in Fig.~1 of the main text, where one can check that the agreement around the maximum is indeed excellent, but that the agreement between the lognormal and the elliptic profiles on the tail is also reasonable.

\ 

\section{Connecting the primordial spectrum to the halo mass function}

The halo mass function determines the number of collapsed halos of a given mass. It is computed from the fraction of collapsed objects $\beta$ as described in Eq. (5) in the main text. 
We can connect the primordial density perturbations to the halo mass with the following procedure.    
From a given dimensionless power spectrum $\Delta(k)$ we can compute the real-space variance
\begin{equation}
    \sigma^2(R)=\int \Delta^2(k)W(k R)\dd \ln k\,,
\end{equation}
where $W(r)=3j_1(r)/r$ is a window function with $j_n$ being the spherical Bessel functions of the first kind. 
For spherical collapse the radius can be linked directly to the enclosed mass via $R=(M/M_*)^{1/3}$, with the reference mass scale $M_*=4\pi \Omega_\mathrm{m}\rho_c/3$ and critical energy density $\rho_c=2.77\times 10^{11} h^{-1}M_\odot / (h^{-1} \mathrm{Mpc})^3$. 
This allows one to compute $\sigma(M)$ at $z=0$. As described in the main text we can compute the variance at any other redshift including the growth factor, i.e. $\sigma(M,z)=\sigma(M,z=0)\cdot D(z)$.

For Gaussian primordial fluctuations the dimensionless power spectrum today is given by  
\begin{equation}
    \Delta_\mathrm{Gauss}^2(k)=\frac{k^3}{2\pi^2}A_s\cdot k^n\cdot T^2 (k)\,,
\end{equation}
where $T(k)$ is the transfer function~\cite{CLASS} approximated by
\begin{equation}
    T(k)\simeq\frac{1}{1+(k/k_*)^{1.6}}
\end{equation}
and $A_s = 2\cdot 10^{-9}$, $n_s = 0.96$ and $k_* = 0.0426/(h^{-1}\mathrm{Mpc})$. The above equations fully determine $\sigma_\mathrm{Gauss}(M,z)$.

\ 

\section{Cluster counts}

The number of halos in a given mass range as a function of redshift is an interesting cosmological observable. From the halo mass function we can compute the number of clusters with masses above a certain threshold, $M_{\mathrm{thr}}$, in a given redshift range, $[z,z+\Delta z]$, as
\begin{equation}
    n_\mathrm{clusters}=\int_{z}^{z+\Delta z}\int_{M_\mathrm{thr}}^\infty\frac{\dd n}{\dd \ln M}\dd \ln M \dd z\, .
\end{equation}
Then, the number of clusters per comoving volume is
\begin{equation}
    N_\mathrm{clusters}=n_\mathrm{clusters}\cdot \Delta V_c\,
\end{equation}
where
\begin{equation}
    \Delta V_c = \int_{z}^{z+\Delta z}\frac{4\pi d_{\mathrm{L}}^2d_H}{(1+z)^2E(z)}\dd z
\end{equation}
with $d_{\mathrm{L}}= (1+z) d_H \int \dd z/E(z)$ being the luminosity distance, $d_H=c/H_0$ the Hubble distance and $E(z)=H(z)/H_0$ the dimensionless rate of expansion.

\bibliography{NGtail}
\end{document}